\documentclass[prl,twocolumn,showpacs,superscriptaddress]{revtex4-1}

\usepackage{amsmath}
\usepackage{amssymb}

\usepackage{longtable}%
\usepackage{dcolumn}
\usepackage{bm}
\usepackage{color}

\usepackage{graphicx}
\usepackage{dcolumn}
\usepackage{bm}
\usepackage{amsmath}
\usepackage{tabu}
\usepackage{multirow}
\usepackage[latin1]{inputenc}
\usepackage{subfigure}
\usepackage{braket}
\usepackage{soul}
\setstcolor{red}
\usepackage{xcolor}
\usepackage{verbatim}
\usepackage[dvipdfm, pdfstartview=FitH, CJKbookmarks=true, bookmarksnumbered=true,
bookmarksopen=true, colorlinks, pdfborder=001, linkcolor=blue, anchorcolor=blue, citecolor=blue]{hyperref}
\newcommand{\sr}[1]{\textcolor{black}{#1}}

\begin{document}
\preprint{Regular article}


\title{Superconducting gap structure of FeSe}

\author{Lin Jiao}
\thanks{These authors contributed equally}
\affiliation{Max Planck Institute for Chemical Physics of Solids,
N\"othnitzer Stra\ss e 40, 01187 Dresden, Germany}
\author{Chien-Lung Huang}
\thanks{These authors contributed equally}
\affiliation{Max Planck Institute for Chemical Physics of Solids,
N\"othnitzer Stra\ss e 40, 01187 Dresden, Germany}
\author{Sahana R\"o{\ss}ler}
\email{roessler@cpfs.mpg.de}
\affiliation{Max Planck Institute for Chemical Physics of Solids,
N\"othnitzer Stra\ss e 40, 01187 Dresden, Germany}
\author{Cevriye~Koz}
\affiliation{Max Planck Institute for Chemical Physics of Solids,
N\"othnitzer Stra\ss e 40, 01187 Dresden, Germany}
%
%
%
\author{Ulrich K. R\"o{\ss}ler}
\affiliation{IFW Dresden, Postfach 270016, 01171 Dresden, Germany}
\author{Ulrich Schwarz}
\affiliation{Max Planck Institute for Chemical Physics of Solids,
N\"othnitzer Stra\ss e 40, 01187 Dresden, Germany}
\author{Steffen Wirth}
\email{wirth@cpfs.mpg.de}
\affiliation{Max Planck Institute for Chemical Physics of Solids,
N\"othnitzer Stra\ss e 40, 01187 Dresden, Germany}

\date{\today}

\begin{abstract}
The microscopic mechanism governing the zero-resistance flow of current in some  iron-based, high-temperature superconducting materials is not well understood up to now. A central issue concerning the investigation of these materials is their superconducting gap symmetry and structure. Here we present a combined study of low-temperature specific heat and scanning tunnelling microscopy measurements on single crystalline FeSe. The results reveal the existence of at least two superconducting gaps which can be represented by a phenomenological two-band model. 
The analysis of the specific heat suggests significant anisotropy in the gap magnitude with deep gap minima. The tunneling spectra display an overall ``U''-shaped gap close to the Fermi level away as well as on top of twin boundaries. These results are compatible with the anisotropic nodeless models describing superconductivity in FeSe. 

\end{abstract}


\maketitle
%
%


\section*{Introduction}

Soon after the discovery of the Fe-based superconductors (Fe-SC) 
great effort has been devoted to unveil their electron paring mechanism. Even after nearly a decade of intensive research,
the symmetry of the superconducting order parameter is still under dispute \cite{Hir2011,Hir2016}, nonetheless most
theories favor an unconventional $s^{\pm}$ symmetry with a sign change of the order parameter
between the hole and the electron Fermi sheets \cite{Maz2008,Kur2008,Wan2009,Kon2010}.

Among the members of the family of Fe-SC,
the binary compound FeSe has attracted considerable attention recently. This is mostly because the
crystal structure of FeSe is regarded as representative
of the entire family of Fe-SC. Further,
the superconducting transition temperature $T_{c}$ $\approx 8 $~K \cite{Hsu2008}
in bulk FeSe can be enhanced up to 37~K by application of
pressure \cite{Miz2008,Med2009,Mar2009,Ima2009} and even to 50 - 100~K
by growing it as a monolayer on a SrTiO$_{3}$ substrate \cite{Yan2012,Ge2015,Fan2015,Zha2016}.
Another unique feature of FeSe is that it undergoes a structural
phase transition from a tetragonal to an orthorhombic phase
at $T_{s} \approx$ 87~K \cite{Mc2009a}, which is not accompanied
or followed by a long-range magnetic order. At $T_{s}$, the $C_{4}$-rotational
symmetry of the underlying electronic system is also spontaneously broken.
The resulting electronic state with a $C_{2}$ symmetry is referred to as a
nematic phase. It is argued that the symmetry of the superconducting
order-parameter should give insight into the collective degree of freedom
that governs both superconductivity as well as nematicity in the Fe-SC \cite{Fer2014}.

However, the situation on the experimental front is far from being resolved.
Even in the case of single crystalline FeSe with relatively simple crystal structure, different experiments indicated different superconducting gap structures. While most experiments detected two superconducting gaps \cite{Kas2008,Dong2009,Kha2010,Pon2011,Lin2011,Abd2013,Kas2014,Hop2016}, no consensus has been reached concerning the magnitude of the superconducting gap as well as on the presence or absence of nodes within the structure. The residual linear component of the thermal conductivity $\kappa_{0}$/T in the $T\rightarrow$ 0 limit, which is particularly sensitive to nodal quasiparticles, revealed contradicting results \cite{Kas2014,Hop2016}. Further, surface sensitive scanning tunnelling spectroscopic (STS) measurements, performed on single crystalline \cite{Kas2014} and thin film samples \cite{Song2011}, detected ``V''-shaped spectra in the superconducting state indicating the presence of nodes. However, STS conducted on the twin boundaries displayed a full gap \cite{Wata2015}, suggesting nodeless superconductivity at the twin boundaries. In order to resolve this issue, it is necessary to perform both bulk and surface sensitive experiments on FeSe. Owing to the marked dependence of the superconducting properties even for FeSe samples grown by the same method \cite{Kas2014,Hop2016},
concerted investigations on identical single crystals are required to establish one of its most fundamental properties, $viz.$, the symmetry of the superconducting order parameter.

Here we report on specific heat $C(T)$ combined with low-temperature ($T \geq$ 0.35~K) scanning tunnelling microscopy (STM) measurements on a stoichiometric FeSe single crystal to establish its superconducting order parameter. As shown below, such a combination of techniques, bulk sensitive $C(T)$ and surface sensitive STM, allows us to unequivocally resolve the superconducting gap structure of FeSe to be nodeless.

\section*{Experimental Results}

\begin{figure}
\centering
	   \includegraphics[clip,width=0.95\columnwidth]{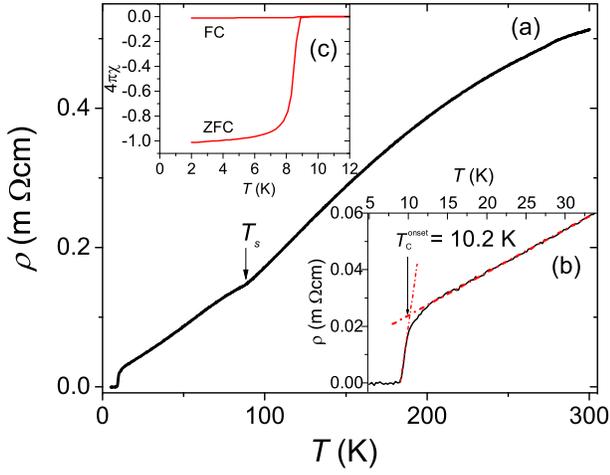}
	\caption{\textbf{Resistivity and magnetization.} (a) Resistivity $\rho (T)$ of FeSe single crystal presented in inset of Fig. 2. The temperature of the structural transition is marked by $T_s$. (b) The same $\rho (T)$ data in (a) zoomed in for $T < 40~K$ showing the superconducting transition. (c) Magnetization measured in a magnetic field of 20~Oe both in zero-field-cooled (ZFC) and field-cooled (FC) protocols.}
	\label{Fig1}
\end{figure}

\begin{figure}
\centering
	    \includegraphics[clip,width=0.95\columnwidth]{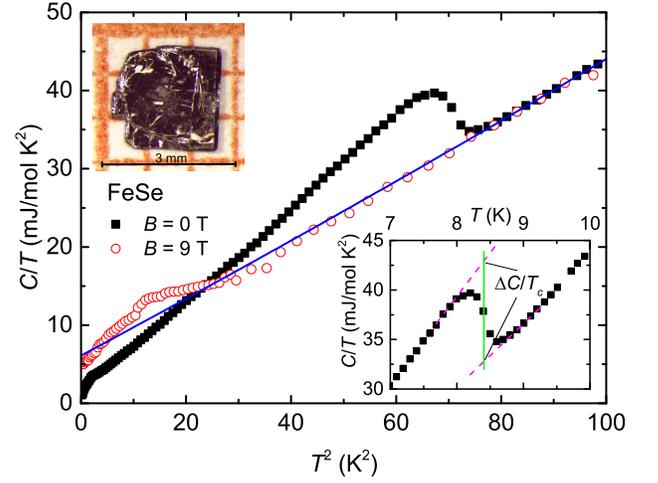}
	\caption{\textbf{Specific heat.} Specific heat divided by temperature, $C/T$ vs $T^{2}$, measured at magnetic fields of zero and 9 T. The solid line represents the normal-state specific heat $C_{n}$. Upper-left inset: a photograph of tetragonal FeSe single crystal used for specific-heat measurements. Lower-right inset: zero-field $C/T$ vs $T$ in an enlarged scale around $T_{c}$. Lines show how $T_{c}$ and $\Delta C/T_{c}$ were determined.}
	\label{Fig2}
\end{figure}

For a general characterization of our single crystal, we measured the temperature dependence of resistivity $\rho(T)$ and magnetization $M(T)$, see Fig.~\ref{Fig1}. These measurements were carried out on the same single crystal which is shown in the inset of Fig.~\ref{Fig2}. In Fig.~\ref{Fig1}(a), $\rho(T)$ measurement along the $[100]_\mathrm{T}$ direction of the tetragonal structure is presented. The resistivities are $\rho_{300\mathrm{K}}$ = 0.51 m$\Omega$ cm at 300 K and $\rho_{15~\mathrm{K}}$ = 0.031 m$\Omega$ cm at 15 K. These values give a residual resistivity ratio (RRR) $\rho_{300\mathrm{K}}$/$\rho_{15\mathrm{K}}$=16.4. The RRR of our crystal is very similar to the FeSe crystal (sample B) investigated by Bourgeois-Hope $et~al.$ in Ref. \onlinecite{Hop2016}. A kink at $T_s$ = 87 K  marks the structural transition temperature. In the inset, Fig.~\ref{Fig1}(b), the onset of superconducting transition can be seen at $T_{c}^{\mathrm{onset}}$=10.2 K. The sample achieves zero resistivity at $T_{c}$ = 8.5 K, which is the superconducting transition temperature of the bulk. In the inset Fig.~\ref{Fig1}(c), $M(T)$ measured in a field of 20 Oe, both in the zero-field-cooled (ZFC) and field-cooled protocol (FC) are shown. The sample displays a full diamagnetic shielding in the superconducting state with $4\pi\chi = -1$.\\

%
\begin{figure}[t]
\centering
	    \includegraphics[clip,width=0.95\columnwidth]{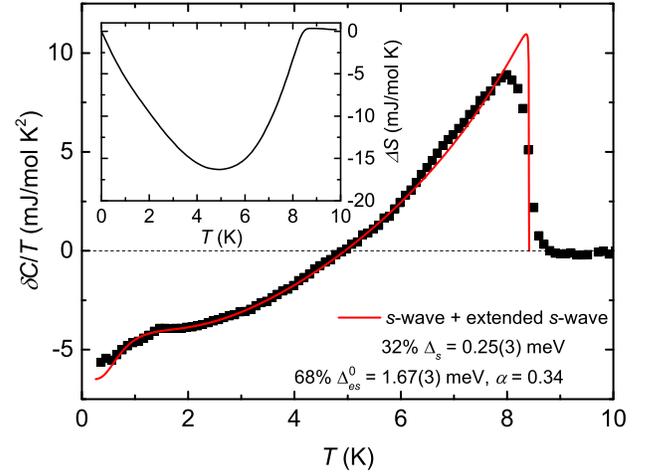}
	\caption{\textbf{Electronic part of the specific heat.} Zero-field electronic specific heat (with the normal-state specific heat being subtracted) divided by temperature. The solid line represents a fit by a smaller $s$-wave plus a larger extended $s$-wave models of the form $\delta C = x~\delta C_s + (1-x)~\delta C_{es}$ with $x=0.32$. The inset shows the entropy conservation required for a second-order phase transition.}
	\label{Fig3}
\end{figure}
\begin{figure*}[t]
\centering
    \includegraphics[clip,width=1.6\columnwidth]{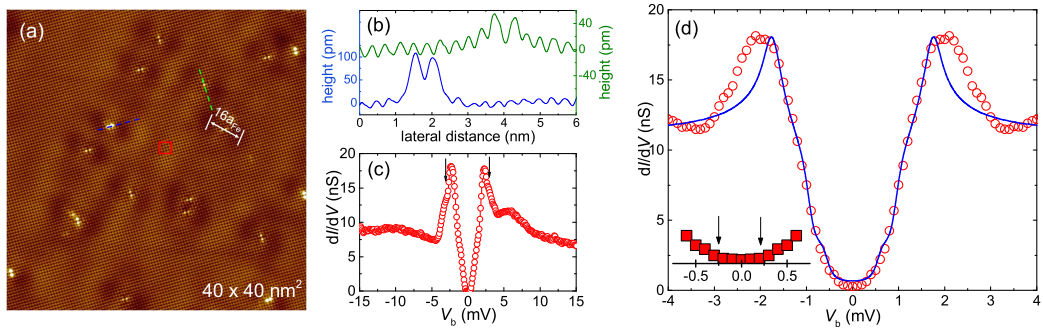}
	\caption{\textbf{STM topography and spectroscopy.} (a) A topography of FeSe on an area of 40 $\times$ 40 nm$^2$ obtained at 0.35 K. 
	The white line mark one of the unidirectional electronic dimer of length $\sim$ 16~$a_{\textrm{Fe}}$, where $a_{\textrm{Fe}}$ is the distance of the Fe-Fe
atoms in the crystal structure. 
The bias voltage and the tunnelling
currents were set at $V_b$ = 10 mV and $I_{sp}$ = 100 pA, respectively. The tunnelling conductance is acquired by the
standard lock-in technique with a small modulation voltage of 0.05 mV$_{\textrm{rms}}$. (b) Line scans along the blue and green lines marked in (a) displaying the heights of the impurities. (c) An average tunnelling spectrum measured within the area of 1 $\times$ 1 nm$^2$ [red square in (a)] at 0.35~K. The arrows indicate ``wing''-like features mentioned in the text. (d) A fit (solid line) of a Dynes gap function to the symmetrized data (open circles) in the $V_b$ range $\pm$3 mV for an $(s+es)$-wave model. For the fit, the thermal broadening as well as the broadening caused by a finite energy resolution was taken into account. Inset :The tunneling conductance at $|V_b|\rightarrow 0$. The arrows mark the voltage range at which d$I$/d$V \approx$ 0 (cf. Supplementary information, Fig.~S11).}
	\label{Fig4}
\end{figure*}
%
%
%
The temperature ($T$) and magnetic field ($B$) dependence of the specific heat $C(T,B)$ was measured on the single crystal imaged in the inset of Fig.~\ref{Fig2}. The zero-field $C/T$ vs $T^{2}$ plot between 0.35 and 10~K presented in Fig.~\ref{Fig2} displays two anomalies, a $\lambda$-like transition at $T_{c} = 8.4(1)$~K and a broad shoulder below 2~K, better seen in Fig.~\ref{Fig3}. This is a typical behaviour of a two-gap superconductor such as MgB$_{2}$ \cite{Bouquet2001}, suggesting the presence of at least two superconducting gaps in FeSe. The $T_{c}$ was determined via local entropy conservation, i.e., the vertical line in the inset of Fig.~\ref{Fig2} segments equal areas in a $C/T$ vs $T$ plot. We describe the normal-state specific heat $C_{n}$ below 10~K by $C_{n}(T) = \gamma_{n}T + C_{lat}(T)$, where $\gamma_{n}T$ is the normal electronic contribution and $C_{lat}(T) = \beta_{3} T^{3} + \beta_{5} T^{5}$ represents the phonon contribution. The fit to $C/T$ is shown as a solid line in Fig.~\ref{Fig2}, which yields $\gamma_{n} = 6.5$~mJ/mol K$^2$, $\beta_{3} = 0.365$~mJ/mol~K$^4$, and $\beta_{5} =$ 1.94$\times 10^{-4}$~mJ/mol~K$^6$. The Debye temperature $\theta_{\texttt{D}}$ calculated from $\beta_{3}$ is 242~K. These parameters are comparable to those reported earlier \cite{Lin2011,Koz2014,Mc2009,Wang2016}. The normalized specific-heat jump at $T_{c}$, $\Delta C/\gamma_{n}T_{c}$, is estimated to be 1.55, which is slightly larger than the weak-coupling value 1.43 of Bardeen-Cooper-Schrieffer (BCS) theory \cite{Bar1957}.
%
%
The excess electronic specific heat contribution in the superconducting state given by $\delta C(T) = C(T, B = 0) - C_{n}(T)$ is plotted in Fig.~\ref{Fig3}. The inset illustrates the satisfaction of entropy conservation $\Delta S = \int_{0}^{T_{c}}(\delta C/T)dT$ justifying the validity of the parameters used to fit $C_{n}(T)$. In the $\delta C/T$ plot, the shoulder below 2~K arising due to the second superconducting gap is clearly visible. To our knowledge, such a shoulder feature has been only reported for pure, polycrystalline samples \cite{Mc2009} with compositions Fe$_{1.01}$Se and Fe$_{1.02}$Se. In order to further examine the superconducting order parameter, the data in Fig.~\ref{Fig3} were fitted to the one-band BCS equation \cite{Bar1957} given by
\begin{eqnarray}
\delta C =
2N(0)\beta k_{\textrm{B}}\frac{1}{4\pi}\int_{0}^{2\pi}d\theta  \nonumber \\
\int_{0}^{\hbar\omega _{\textrm{D}}} [ -\frac{\partial f}{\partial E}\left ( E^{2}+\frac{1}{2}\beta \frac{d\Delta^{2}(T,\theta)}{d\beta} \right ) ]d\varepsilon-\gamma_{n}T,
\end{eqnarray}
%
%
where $N(0)$ is the density of states at the Fermi surface, $\beta =$ 1/$k_{\textrm{B}}T$, $E = [\varepsilon^{2}+\Delta^{2}(T, \theta)]^{1/2}$, $f = (1 + e^{\beta E})^{-1}$, and $\Delta(T, \theta) = \Delta_{es}^0(T)(1 + \alpha~\mathrm{cos} 4\theta)$ an extended $s$-wave where $\alpha$ and $\theta$ represent the gap anisotropy and polar angle, respectively \cite{Maier2009,Chubukov2009}. Note that in an angle-integrated measurement, the functional forms containing $\mathrm{cos} 4\theta$ and $\mathrm{cos} 2\theta$ result in the same parameters and hence, it is not possible to distinguish between an extended $s$-wave and a $d$-wave symmetry. We found that a single anisotropic $s$-wave model, either with or without accidental nodes, does not fit to the $\delta C/T$ data (see Supplementary information, Figs. S1-S3.). 
Since $\delta C$ indicated a signature of two superconducting gaps, we also tried a phenomenological two-gap model \cite{Suh1959,Mos1959,Zeh2013}
by taking a sum of either two $s$-wave-gaps (see Supplementary information, Fig. S4) or an $s$-wave + an extended $s$-wave gaps ($s+es$), Fig.~\ref{Fig3} (see also Supplementary information, Fig. S5), 
to describe the data. 
In the fitting, more weight was given to the low-temperature data, i.e., the data below $T =$ 5 K. 
The reason being, close to $T_{c}$ the thermal fluctuations become stronger, and may result in increasing deviation of the data from the applied models which are based on a mean-field framework.
We found that both models lead to satisfactory fits.
Hence, the exact superconducting gap structure of FeSe cannot be unambiguously determined from analyzing the specific-heat data alone. However, as discussed later, with the help of fitting several models also to the tunnelling spectra (see Supplementary information, Figs. S8-S11), we could select ($s+es$) model as a proper model to describe the $C(T)$-data, with the larger gap assigned to the extended $s$-wave. The goodness of fit for each model tried here is presented in the Supplementary information Table S1. 
The gap values in the $T \rightarrow 0 $ limit estimated from the ($s+es$) model fitting are: a small $s$-wave gap of $\Delta_{s}(0) = 0.25(3)$ meV and a large anisotropic extended $s$-wave gap of $\Delta_{es}^0(0) = 1.67(3)$ meV with $\alpha = 0.34$. The value of $\alpha <$ 1 obtained here clearly rules out the presence of accidental nodes \cite{Hin2015}. \sr{Since the obtained isotropic gap value $\Delta_{s}(0)$ is very small, a possible anisotropy of this gap would be beyond the resolution of our experiments. Further, a recent heat capacity study of FeSe single crystals by Wang $et~al.$ \cite{Wang2016} reports a small residual value of the electronic specific heat originating from low-energy quasiparticle excitations indicating either line nodes or deep gap minima. However, their experiment did not show the shoulder in $C(T)$ which we observe below 2~K and interpret as the mark of a second, smaller superconducting gap. In our case, this shoulder limits the analysis of the functional form of $\delta C(T)$ as $T\rightarrow 0$. Therefore, if we consider the data only below 1~K, the presence of accidental nodes can not be ruled out based on the specific heat analysis.} 

\begin{figure*}[t]
\centering
\includegraphics[clip,width=1.6\columnwidth]{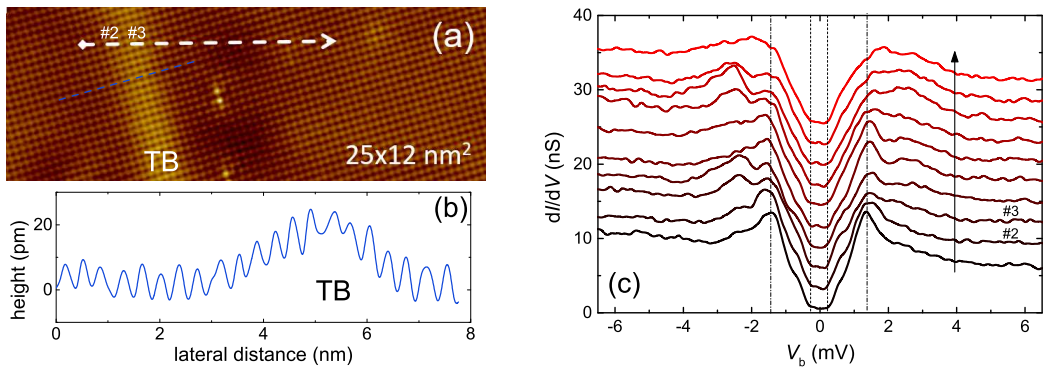}
\caption{\textbf{Tunneling spectra at a twin boundary.} (a) A 25$\times$12 nm$^2$ topography of FeSe with a twin boundary (TB), which is zoomed from Fig. S6 (Supplementary information). (b) A line scan along the blue line depicted in (a). (c) Ten tunnelling spectra measured laterally at equidistant positions along the white dashed line in (a). The black arrow in (c) represents the direction of measurement shown in (a). Spectra $\#$2 and $\#$3 are measured on the TB. Curves are equally shifted vertically for clarity. The bias voltage and the tunnelling currents were set at $V_b$ = 10 mV and $I_{sp}$ = 100 pA, respectively.}
\label{Fig5}
\end{figure*}
In order to determine the superconducting gap
structure of FeSe microscopically, we performed STM measurements at 0.35~K.
The topographic images, $e. g.$ Fig.~\ref{Fig4}(a),
revealed atomically resolved clean Se-terminated surfaces indicating good sample quality, very similar to our previous report \cite{Ros2015}. The Se-Se
distance $a_{Se-Se} = 3.7(1)$ \AA~observed here is in line with the distance
of 3.7702(5)~\AA~found by x-ray diffraction on our crystals \cite{Koz2014}.
A few protrusions (see Fig.~\ref{Fig4}(b) for line scans across the impurities) on the top of the surface likely correspond to 
Se-Se bound atoms left over from the top-most counter layer while cleaving, or to an impurity atom occupying the Fe-site underneath the topmost Se-layer. 
Alternatively, recent density functional theory (DFT) based calculations suggested that Fe-site vacancies can perturb orbitals on neighbouring Se-sites, thereby producing atomic dumbbells \cite{Den2016}. All these defects act as impurities and induce additional dumbbell-like unidirectional depressions in the topography, known as ``electronic dimers'' \cite{Song}, marked by
the white line in Fig.~\ref{Fig4}(a). Interestingly, the unidirectional behaviour of these
 electronic dimers can be unveiled from the fact that the orientation of the
 dimers is independent of the orientation of the impurities but 
 rotates by 90$^\circ$ across the twin boundary (TB) obvious from the bright stripe in the topography, Supplementary information, Fig.~S6. This behaviour represents a broken $C_4$ symmetry in the impurity scattering. The length of the electronic dimers is $\approx$ 16~$a_{\textrm{Fe}}$, where $a_{\textrm{Fe}}$ is the distance of the Fe-Fe
 atoms in the crystal structure, which is consistent with a previous report \cite{Song}. We note that all electronic dimers are oriented in the same direction in Fig.~\ref{Fig4}(a) indicating the entire image consists of a single crystallographic domain.\\

In Fig.~\ref{Fig4}(c), a tunnelling spectrum averaged over an area marked in Fig.~\ref{Fig4}(a) is presented. Since the spectra were measured approximately in the middle of Fig.~\ref{Fig4}(a), a possible TB is at least a minimum distance of 20 nm away.
The most prominent feature of the spectrum is that, as $V_b \rightarrow 0$, the spectrum attains a ``U''-shape. Here, ``U''-shape refers to the finite energy range within which the experimental tunnelling conductance is zero, as more clearly seen in Fig.~\ref{Fig4}(d). Due to the estimated small magnitude of the smaller gap (see above) a zero tunnelling conductance is only expected within a very narrow energy range. A tunnelling conductance of zero indicates the absence of quasiparticle excitations within the superconducting gap, thus providing strong evidence for a nodeless superconductivity in FeSe.
However, the spectrum contains additional hump-like features at energies $\approx -$10 mV and +5.4 mV. These represent either simply the bottom of an electron band and the top of a hole band, respectively, or more complex phenomena such as a density-wave type ordering \cite{Ros2015} or an electron-boson coupling \cite{Song2014}. In addition, there are ``wing''-like features contained in the coherence peaks at energies $V_b \approx \pm2.75$~mV, indicated by black arrows in Fig.~\ref{Fig4}(c). These features may be related to the fine details of the band structure, such as spin-orbit coupling \cite{Bor2015} induced band-hybridization \cite{Sch2013,Hin2015,Kom2015}, which is beyond the scope of this paper.

In an attempt to describe the tunnelling spectra, we used a Dynes gap function \cite{Dynes1978} to fit the data. Within the standard picture of the BCS model \cite{Bar1957}, the tunnelling conductance $\mathrm{d}I/\mathrm{d}V \propto f_D(E) $, where
\begin{equation}
\label{fDyne}
f_D=\int_{0}^{2\,\pi}d\,\theta\;\mathrm{Re}\,\frac{e\,V-\mathrm{i}\,\varGamma}{\sqrt{(e\,V-\mathrm{i}\,\varGamma)^2-[\Delta(\theta)]^2}}~.
\end{equation}
Here, the spectral broadening is given by the inverse quasiparticle lifetime $\varGamma$. 
Since the experimental spectra did not indicate the presence of nodes, we first tried a single extended $s$-wave gap function $\Delta(\theta)=\Delta_{es}^0(1 + \alpha~\mathrm{cos} 4\theta)$ to fit the data (see Supplementary information, Fig. 11). Although the fitted curve followed the experimental data well in the applied bias voltage range $0.5~\mathrm{meV} < |V_b| < 0.75~$meV, below 0.5 meV, the fitted curve deviated from the experimental data as $V_b \rightarrow 0$. This suggested the presence of a small second gap as already inferred from the specific heat analysis. However, the coherence peaks corresponding to the smaller gap could not be resolved in our experiments due to its small magnitude, which is at the limits of our instrumental resolution. To take this into account, we
included an energy resolution of 0.16 meV in the fit procedure, which accounts for the spectral broadening caused by thermal effects (0.35 K) as well as a finite modulation voltage (0.05 mV$_\mathrm{rms}$).
In Fig.~\ref{Fig4}(d), the best fit to the experimental spectrum is presented. This fit corresponds to an ($s+es$) model with an $s$-wave gap of $\Delta_{s}=$0.6(1) meV and an extended $s$-wave gap $\Delta_{es}^0 = 1.35(2)$ meV and $\alpha = 0.30(1)$. These values are slightly different than those obtained from the specific heat analysis. Here we would like to emphasize that the models considered here should not be taken exhaustive, rather it should be understood as a minimum ansatz to describe the overall behaviour of the spectrum, which agrees semi-quantitatively with the specific-heat analysis. By considering the raw data alone and leaving the models aside, the multigap nature of the superconducting gap is derived from the specific heat measurements, whereas, the nodeless nature of the gap is concluded from the tunnelling spectroscopy measurements.\\

Following this indication towards nodeless superconductivity in FeSe, we now show --- using STM/STS --- that the gap structure appears to remain nodeless on different crystallographic twin domains as well as at the TB. In Fig.~\ref{Fig5}(a), an STM topographic image over an area of 25$\times$12 nm$^2$ containing a TB is presented. This image is a part of the topography of 40 $\times$40 nm$^2$ presented in Fig. S6 (Supplementary information). A height scan across the TB is shown in Fig. \ref{Fig5}(b). Several spectra were measured along the white line in Fig.~\ref{Fig5}(a) in such a way that the spectra were distributed on either side as well as on the TB. As can be seen in Fig.~\ref{Fig5}(c), the spectra retains an overall ``U''-shape across the TB, warranting the robustness of the nodeless gap structure in our single crystal. Alternatively, Watashige $et.al$ \cite{Wata2015} observed a lifting of nodes in the vicinity of a TB and interpreted this finding in terms of time reversal symmetry breaking caused by a $\pi/2$ rotation of the crystallographic domains. They also found that the influence of the TB on the superconducting gap structure extends up to a length scale of more than 50 nm. Since the spectra shown in Fig.~\ref{Fig5}(c) were measured only up to 9 nm away from the twin boundary, our results shown in Fig.~\ref{Fig5} do not directly contradict those of Ref. [\onlinecite{Wata2015}]. However, we did not find any signature of pair-breaking by observing a formation of bound states in the spectra taken on the TB, which was suggestive of a time reversal symmetry breaking. \sr{For the sake of confirming the U-shape of the tunneling spectra at small $V_b$ as a common feature of our sample, we performed STM/STS on a second crystal. In this case we conducted our measurements on an area of 100 $\times$100 nm$^2$ without any TB, see Supplementary information Fig. S7. As can be seen in Fig. S7(b), even the small gap could be resolved in some cases in the tunneling spectra. However, within a small range of $V_b $, the spectra retain a U-shape indicating the absence of low energy quasiparticle excitations owing to finite superconducting gap over the Fermi surface.}


\section*{Discussion}
In the framework of a single-band BCS theory, the zero-temperature upper critical field $H_{c2}(0)$ is proportional to $(\Delta/v_{\texttt{F}})^{2}$, $v_{\texttt{F}}$ being the Fermi velocity \cite{Shulga2002}. In a phenomenological two-gap model, $H_{c2}(0)$ is set by the larger gap, and the critical field for the smaller gap $H^{*}(0)$ can be determined by thermodynamic, e.g., specific-heat \cite{Bouquet2002} and thermal-conductivity \cite{Hop2016} measurements. Shubnikov-de Haas oscillation measurements have reported comparable values of $v_{\texttt{F}}$ for the different Fermi sheets in FeSe \cite{Tera2014}. By taking the maximum gap value $\Delta^\mathrm{max}_{es} = \Delta_{es}^0(1+\alpha)$ and $\Delta_{s}(0)$ obtained from the $C(T)$ analysis, we estimate, $H^{*}(0)/H_{c2}(0) = [\Delta_{s}(0)/\Delta^\mathrm{max}_{es}(0)]^{2} = (0.25/2.23)^{2} \approx 0.01$. This value is in good agreement with the data reported by Bourgeois-Hope $et$ $al.$~\cite{Hop2016}, thus further supporting the validity of the current analysis. As far as the absence of nodes is concerned, our results are also in good agreement with recent thermal conductivity \cite{Hop2016}, penetration depth \cite{Tek2016}, and microwave conductivity \cite{Li2016} measurements on single crystals of similar quality. \\ 

It is worthwhile to discuss the possible origin of the discrepancy between the STM results presented here in comparison to those in Refs. [\onlinecite{Kas2014,Wata2015}]. The nodes observed in FeSe are considered accidental, i.e., they are not imposed by symmetry \cite{Hir2011,Hir2016}. Theoretical investigations of multiorbital microscopic models have suggested that the nodes in the Fermi surface can be lifted by disorder \cite{Mis2009} or external strain \cite{Kan2014}. One of the ways to get a semi-quantitative estimation of the degree of disorder in a sample is to look at its RRR value.
However, we would like to point out that in the particular case of FeSe, the RRR calculated by taking the resistivity values from above and below $T_{s}$ contain additional contributions other than initial intrinsic disorder of the crystal which exists at room temperature. As observed by Kn\"oner $et~al.$ \cite{Kno2015}, cooling the samples through $T_{s}$ induces different twin states in the samples in question; which together with the finite in-plane anisotropy can produce different resistivity values below $T_{s}$. A similar observation was also made in Ref. \onlinecite{Hop2016}. Therefore, the crystals showing lower RRR likely contain more twins, and TB are considered accountable for lifting the nodes \cite{Wata2015}.  Nonetheless, a very recent thermal conductivity \cite{Hop2016} measurement on samples grown by flux-vapour transport \cite{Boe2013} with the RRR values similar to those used in Refs. [\onlinecite{Kas2014,Wata2015}], exhibited two-gap nodeless superconductivity. It is rather intriguing that such negligibly small differences in the samples appear to be sufficient to influence the superconducting gap structure in FeSe. \\

Our observation of two superconducting gaps $\Delta_{s}$ and $\Delta_{es}$ with strongly different gap magnitudes, $i.e.$ $\Delta_{s} \ll \Delta_{es}$, indicates that superconductivity appears presumably in one band (producing a large gap $\Delta_{es}$) and may induce a second small gap $\Delta_{s}$ in another band due to a proximity effect \cite{Sch2013,Mc1968}. Nonetheless, both gaps open at the same temperature, but may have different temperature dependencies \cite{Hir2011}. So far in FeSe, only one hole Fermi sheet and one electron Fermi pocket are detected by angle-resolved photoemission spectroscopy (ARPES) \cite{Mal2014} and quantum oscillation experiments \cite{Tera2014,Wat2015a}. If this is correct, then an inter-band extended $s$-wave pairing with a sign reversal of the order parameter between different Fermi surface sheets might be the likely mechanism of superconductivity in FeSe \cite{Maz2008}. However, there are experimental indications for more than one electron pocket crossing the Fermi energy \cite{Wat2015, Huy2014}. In this case, a more exotic pairing mechanism such as band-hybridization induced odd frequency pairing can be expected \cite{Sch2013}.\\


In summary, we have investigated the superconducting gap structure of FeSe in a combined study of scanning tunnelling microscopy and specific heat measurements. \sr{The results indicate
multigap superconductivity in FeSe single crystals.} Our analysis suggests that the gap is of ($s+es$) type. The isotropic $s$-wave gap is much smaller than the anisotropic $s$-wave gap. Additionally, the tunnelling spectroscopy indicate at a superconducting gap which remains nodeless also on twin boundaries. These experimental results are expected to provide important ingredients for a unified theory of the superconducting paring mechanism for all FeSe-related superconductors.\\

\textit{Note~added}: In the revising stage of this manuscript we became aware of a new STM study on FeSe reported very recently \cite{Spr2016}. Our conclusions presented here are in excellent agreement with these complementary investigations in which the Bogoliubov quasipaticle scattering interference (BQPI) was used to determine the superconducting gap symmetry as extremely anisotropic, but nodeless with an OP changing sign between the hole and electron pockets. In addition, Ref. [\onlinecite{Spr2016}] also provides evidence for an orbital-selective Cooper pairing in FeSe.



\section*{Methods}

The single crystals were grown by chemical vapour transport\cite{Koz2014,Ros2016} of stoichiometric FeSe powder containing $\alpha$-Fe of less than 300 ppm. The ratio of FeSe to the transport additive AlCl$_3$ was taken as 50:1. Typically, a mixture of 1 g of FeSe powder and 20 mg of AlCl$_3$ was placed in a quartz ampoule of length 10 cm and diameter 2 cm prepared inside an argon-filled glove box. The ampoule containing the mixture was evacuated, sealed, and placed horizontally inside a two-zone furnace at temperatures from $T_{2}$=673~K and $T_{1}$=573~K. The crystal growth was carried out for 2 months. Finally, the ampoule was quenched in water. The product, which contained plate-shaped single crystals with edge lengths up to 400 $\mu$m perpendicular to the $c$ axis, was washed repeatedly in ethanol to remove remaining condensed gas phase, dried under vacuum and stored in the glove box. By extending the growth time to one year, larger single
crystals with dimensions up to $4 \times 2 \times 0.03$ mm$^3$ could be grown. The specific heat $C(T,B)$ was measured down to 0.5~K using a thermal-relaxation method in a physical property measurement system (Quantum Design) with the magnetic field $B$ applied parallel to the [001] direction of the single crystal. The scanning tunnelling microscopy/spectroscopy measurements were performed in an ultra-high vacuum ($p < 3 \times 10^{-9}$ Pa) cryogenic STM with a base temperature $T\approx 0.35$~K. The bias voltage and the tunnelling
currents were set at $V_b$ = 10 mV and $I_{sp}$ = 100 pA, respectively. The tunnelling conductance is acquired by the standard lock-in technique with a small modulation voltage of 0.05 mV$_{\textrm{rms}}$. Given the total energy resolution $\Delta E$ of the STM is limited by $\Delta E \approx \sqrt{(3.5k_{\mathrm{B}}T)^2+(2.5 eV_{\mathrm{mod}})^2}$ \cite{Lou2001}, the low temperature (0.35 K) and the small modulation voltage (0.05 mV$_{\textrm{rms}}$) used here allows us to resolve the fine structure of the superconducting gap. The FeSe single crystals were cleaved $in~situ$ at 20~K before being inserted into the STM-head.


%

%
%

\section*{Acknowledgements}

Financial support from the Deutsche Forschungsgemeinschaft (DFG) within the Schwerpunktprogramm SPP1458 is gratefully acknowledged. L.J. thanks the Alexander-von-Humboldt foundation for financial support. C.L.H. acknowledges technical support from C. Klausnitzer. We thank A.~Akbari, S. Borisenko, Y. Grin, M. Haverkort, P. J. Hirschfeld, P.~Thalmeier, and L.~H. Tjeng for discussions.

\end{document}